\documentclass[journal]{IEEEtran}
\usepackage{amsmath,amsfonts}
\usepackage{algorithmic}
\usepackage{algorithm}
\usepackage{array}
\usepackage[caption=false,font=normalsize,labelfont=sf,textfont=sf]{subfig}
\usepackage{textcomp}
\usepackage{stfloats}
\usepackage{url}
\usepackage{verbatim}
\usepackage{graphicx}
\usepackage{cite}
\usepackage{soul}

\usepackage{xcolor}
\newcommand{\gp}[1]{{\color{black}{#1}}}

\begin{document}

%\title{Unleashing the Power of Wireless Human-Machine Collaboration in 6G Networks}
%\title{Toward Wireless Human-Machine Collaboration in 6G Networks: Vision, Requirements, Technologies, and Challenges}
%\title{Wireless Human-Machine Collaboration Systems in the 6G Era}
\title{Toward Wireless Human-Machine Collaboration in the 6G Era}

\author{Gaoyang Pang, Wanchun Liu,~\IEEEmembership{Senior Member,~IEEE,} 
Chentao Yue,
%Zhibo Pang,~\IEEEmembership{Senior Member,~IEEE,} 
%Dusit Niyato,~\IEEEmembership{Fellow,~IEEE,}
Daniel E. Quevedo,~\IEEEmembership{Fellow,~IEEE,}\par
Karl H. Johansson,~\IEEEmembership{Fellow,~IEEE,}
Branka Vucetic,~\IEEEmembership{Life Fellow,~IEEE,} and Yonghui Li,~\IEEEmembership{Fellow,~IEEE}
%\vspace{-1cm}
        % <-this % stops a space
%\thanks{The work of W. Liu was supported by the Australian Research Council’s Discovery Early Career Researcher Award (DECRA) Project DE230100016. \textit{(Corresponding author: W. Liu.)}}% <-this % stops a space
\thanks{G. Pang, W. Liu, C. Yue, D. Quevedo, B. Vucetic, and Y. Li are with the School of Electrical and Computer Engineering, The University of Sydney, Sydney, NSW 2006, Australia (e-mail: \{gaoyang.pang; wanchun.liu; chentao.yue; daniel.quevedo; branka.vucetic; yonghui.li\}@sydney.edu.au).}
\thanks{K. H. Johansson is with the Division of Decision and Control Systems, School of Electrical Engineering and Computer Science, and Digital Futures, KTH Royal Institute of Technology, 100 44 Stockholm, Sweden (e-mail: kallej@kth.se).}
%\thanks{Z. Pang is with the Department of Automation Technology, ABB Corporate Research, 72178 Vasteras, Sweden, and also with the Department of Intelligent Systems, Royal Institute of Technology, 11758 Stockholm, Sweden (e-mail: pang.zhibo@se.abb.com; zhibo@kth.se).}
%\thanks{D. Niyato is with the School of College of Computing and Data Science, Nanyang Technological University, Singapore 639798, (e-mail: dniyato@ntu.edu.sg)}
}

%\markboth{IEEE Communications Magazine}%
%{G. Pang \MakeLowercase{\textit{et al.}}: Unleashing the Power of Wireless Human-Machine Collaboration in 6G Networks}


\maketitle

\begin{abstract}
The next industrial revolution, Industry 5.0, will be driven by advanced technologies that foster human-machine collaboration (HMC). It will leverage human creativity, judgment, and dexterity with the machine’s strength, precision, and speed to improve productivity, quality of life, and sustainability. Wireless communications, empowered by the emerging capabilities of sixth-generation (6G) wireless networks, will play a central role in enabling flexible, scalable, and low-cost deployment of geographically distributed HMC systems. In this article, we first introduce the generic architecture and key components of wireless HMC (WHMC). We then present the network topologies of WHMC and highlight impactful applications across various industry sectors. Driven by the prospective applications, we elaborate on new performance metrics that researchers and practitioners may consider during the exploration and implementation of WHMC and discuss new design methodologies. We then summarize the communication requirements and review promising state-of-the-art technologies that can support WHMC. Finally, we present a proof-of-concept case study and identify several open challenges.
\end{abstract}

%\begin{IEEEkeywords}
%Human-machine collaboration, Human-cyber-physical systems, Industry 5.0, Wireless Communications.
%\end{IEEEkeywords}

%\vspace{-0.8cm}
\section{Introduction}
\IEEEPARstart{T}{he} next industrial revolution, Industry 5.0, envisions a future of unprecedented human-machine collaboration (HMC) \cite{ref1}, where increasingly powerful and precise machines work in harmony with the unique creativity of humans.
Unlike previous industrial paradigms, Industry 5.0 places humans back at the center of the production process, harnessing their creativity, intuition, and problem-solving skills while delegating repetitive, monotonous, hazardous, and heavy-duty tasks to machines. This division of labor enables humans to focus on more stimulating and value-adding activities that are difficult to automate. In this paradigm, humans act as guides, quality assurance experts, and decision-makers, collaborating with machines either by teaching and cooperating side by side or by remotely controlling an actuator as an extension of their own body. To enable timely and reliable collaborations between geographically distributed humans and machines, wireless HMC (WHMC) has emerged as a critical enabler, providing the required flexibility and scalability. 

\gp{The fifth generation (5G) network is enabling the “factory of the future” through digitization and HMC~\cite{ref0}.}
Nevertheless, WHMC applications in different vertical industries have unique generic requirements that go beyond ultra-reliable and low-latency communication (URLLC), such as global connectivity, high mobility, and low jitter. These requirements have not been fully met in 5G networks, whose capabilities remain insufficient to realize the full vision of WHMC. The sixth generation (6G) wireless network, however, will introduce advanced networking technologies and mark a new era of connectivity and communication capabilities \cite{ref2}. By seamlessly integrating terrestrial and non-terrestrial networks, 6G will extend accessibility for both humans and machines, facilitating remote control and monitoring of dynamic physical processes. Moreover, 6G will support the transformation of human-machine interfaces (HMIs), allowing real-time, high-bandwidth, multi-modal data exchange between humans and machines. This allows the encoding of human intuitive control and provides the remote human operator with an immersive experience during collaboration. Furthermore, artificial intelligence (AI) will be seamlessly embedded into 6G, empowering machine intelligence to unlock the full potential of WHMC. 

Despite the promise of 6G, reliable WHMC cannot be simply achieved by connecting humans, machines, and networks without considering their interdependencies~\cite{ref3_1}. It requires a new networked system topology with coupled wireless human and automated control loops. Thus, WHMC lies at the intersection of several engineering and scientific disciplines, such as human psychophysics, behavioral economics, mechanical design, communication, and control engineering. Conventional siloed research efforts in these fields are not adequate since the dynamics and interdependencies among human, communication, and control systems demand a more holistic approach. Fully harnessing the potential of WHMC requires the convergence of these fields, which presents formidable challenges. Key barriers include the development of new fundamental mathematical models, performance metrics, and design technologies consistent with a unified WHMC architecture. 

In this paper, we take an initial step toward addressing these challenges, aiming to bridge existing gaps for the large-scale rollout of WHMC. We aim to motivate researchers to treat the wireless, human, and control systems as a united system, enabling fully integrated designs supporting a wide range of emerging applications. 
Specifically, we propose a general framework of WHMC to establish a comprehensive understanding of the concept.
We develop new design methodologies tailored to WHMC systems, addressing their unique challenges and constraints.
We then identify and discuss key design requirements and state-of-the-art technologies, providing researchers with insights into the evolving WHMC research landscape. 
We also present a proof-of-concept case study to demonstrate and validate the effectiveness of the proposed WHMC framework.
Finally, we highlight key open challenges in WHMC, as well as broader issues extending beyond technology.

%\vspace{-0.3cm}
\section{Envisioned WHMC}
\subsection{WHMC and Its Core Components}\label{sec:WHMC}
\begin{figure}[!t]
\centering
%\color{blue}
\includegraphics[width=3.5in]{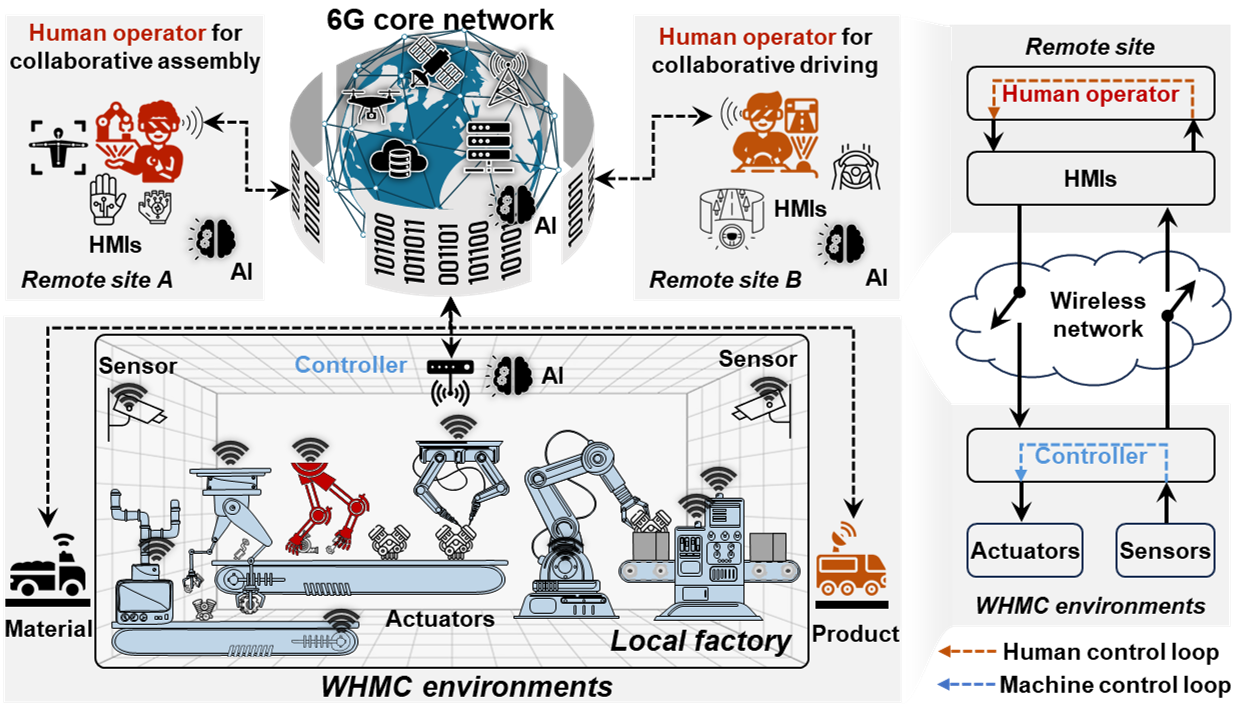}
%\vspace{-0.7cm}
\caption{Illustration of WHMC in advanced manufacturing with abstracted key components.}
\label{fig_1}
%\vspace{-0.0cm}
\end{figure}

An illustrative example of WHMC in advanced manufacturing is collaborative engine assembly, as shown in Fig.~\ref{fig_1}. In such a scenario, actuators, such as robot arms, execute the instructions of the human operator in a remote site (Site A), and the machine control commands generated by a centralized automated controller within the factory. Similarly, in collaborative driving for logistics operations outside the factory, vehicle actuators follow the maneuvering intentions of a remote human operator (Site B) while concurrently responding to steering and motion control commands issued by on-board autonomous controllers. These examples reveal the following core components of WHMC.

\textit{1) Human:} Remote operators make critical decisions and execute collaborative tasks using their creativity, judgment, and dexterity. 

\textit{2) Interface:} HMIs deployed at remote sites capture human instructions and replicate the HMC environment, enabling the human operator to both perform and monitor the collaboration task effectively. 

\textit{3) Sensor:} Distributed sensors in the HMC environment (e.g., factory floor) capture real-time state information of the WHMC system, including variables to be controlled (e.g., positions), the status of actuators, task execution progress, and environmental dynamics. 

\textit{4) Actuator:} Actuators are distributed in the HMC environment to implement control actions issued by both the remote human operator and the autonomous machine controller. 

\textit{5) Controller:} This component provides the computational control logic for autonomous actuators. It integrates sensor data to generate machine control commands that ensure the collaboration goals are achieved. 

\textit{6) Network:} The communication network bridges the remote site and the HMC environment, ensuring reliable and timely exchange of data and control signals between the two domains.

Together, these core components form the generic WHMC architecture, which can be flexibly adapted to diverse Industry 5.0 applications.

%\vspace{-0.3cm}
\subsection{Network Topologies of WHMC} \label{sec: network topologies}
In WHMC, the human operator, the machine controller, and the task are the three critical entities. Their wireless connection topologies are important in three aspects.
\begin{figure}[!t]
\centering
%\color{blue}
\includegraphics[width=3.5in]{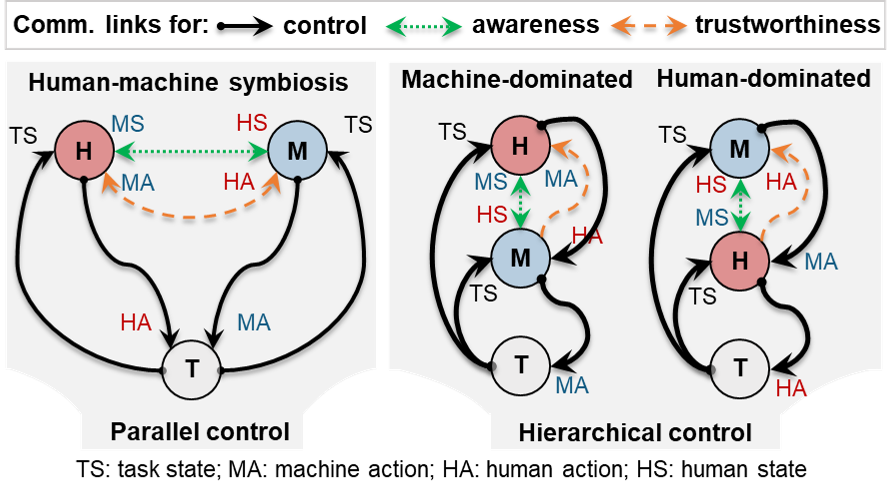}
%\vspace{-0.7cm}
\caption{WHMC topology at a glance of agent autonomy between human H, and machine M, for a collaboration task T.}
\label{fig_2}
%\vspace{-0.0cm}
\end{figure}

\subsubsection{Agent autonomy}
Agent autonomy describes the roles of a human operator and a machine controller in a given collaboration task, as shown in Fig.~\ref{fig_2}. In a human-machine symbiosis system, both agents’ actions are applied to the actuators simultaneously. In the machine-dominated system, the machine controller is supervised by the human operator, and only machine actions drive the actuators. In the human-dominated system, the human operator makes decisions based on the recommendation from the machine controller, and only human actions drive the actuators. Ignorance, awareness, and trustworthiness are defined to describe the information available to these two agents for decision-making. Without information exchange (Ignorance), the agents ignore each other. Awareness allows the observation of the state of their counterpart. When both the state and action of the counterpart are known, an agent can estimate its trust in collaboration.

\subsubsection{Agent multiplicity}\label{sec:agentMulti}

Analogous to multiple channel access, agent multiplicity describes human-actuator associations in multi-agent WHMC systems. In orthogonal human control, each human operator manages a different collaboration task. In non-orthogonal human control, at least one collaboration task is shared by multiple human operators who are not permitted to handle multiple tasks simultaneously. Non-orthogonal human coordination control describes the case when the human operator is able to coordinate multiple tasks concurrently. Considering the capabilities of the machine agent in fast response and multi-tasking, the machine controller is assumed to be centralized and always capable of handling multiple tasks in the above three cases. 

\subsubsection{Agent geography}
Agent geography describes the agent-task distance of various WHMC systems across different altitudes. Short-distance WHMC systems within a single wireless coverage area can be easily supported by advanced wireless technology. 
%It will become more challenging for geographically distributed mid- and long-distance WHMC systems, whereby transmissions may need to traverse beyond the wireless access segment and even metro access segments.
\gp{For mid- and long-distance WHMC, packets often traverse fronthaul/midhaul/backhaul (x-haul) segments and metro/core transport, so end-to-end latency and jitter are shaped by transport switching, queuing, and path diversity, not only by the air interface. This makes transport design a first-order factor in sustaining stable WHMC across sites.}

%\vspace{-0.3cm}
\subsection{Applications Empowered by WHMC}
WHMC holds immense promise across various domains.
One major area is advanced manufacturing as presented in Section~\ref{sec:WHMC}. Besides, WHMC also enables synergistic interactions in following applications.

\subsubsection{Agriculture}
Drone-assisted crop harvesting is a representative application of WHMC. Ground harvesting robots face challenges in safely grasping the objects randomly distributed in three-dimensional (3D) agricultural environments due to blind spots. Grasping invisible objects requires cooperation with external devices that provide the global position feedback of objects. WHMC enables a human operator to control a drone to deliver a bird’s eye view for object localization to close the vision gap of ground robots.

\subsubsection{Transportation}
In autonomous driving, human assistance remains vital in unseen situations or challenging cases. A remote human operator may support the self-driving vehicle’s decisions at an intersection. Through a driving-simulation HMI, the operator changes the vehicle’s steering direction based on the dynamic traffic conditions. The local machine controller inside the vehicle maintains a minimal safe distance to pedestrians and other vehicles.

\subsubsection{Healthcare}
Telesurgery is a cutting-edge application of WHMC, which enables highly qualified and experienced surgeons to perform critical surgeries remotely. The surgeons’ actions are captured and transmitted by HMIs, and then used to control a robotic device in the operating room to replicate the surgeons’ operations. The local machine controller inside the operating room provides assistive operations, such as bleeding control, blood/oxygen circulation, and handover of surgical instruments.

% \begin{figure}[!t]
% \centering
% \includegraphics[width=3.5in]{Fig3.png}
% %\vspace{-0.7cm}
% \caption{The potential application fields of WHMC  with three representative scenarios in agriculture,  transportation, and healthcare.}
% \label{fig_3}
% %\vspace{-0.0cm}
% \end{figure}

%\vspace{-0.3cm}
\subsection{Performance Metrics of WHMC}
In conventional cyber-physical control systems (CPCSs), the performance is characterized by the control cost, which is defined as a function of plant state and control inputs. The cost function design depends on the specific goals and constraints of the CPCS. For example, in production-line automation, the cost function quantifies mission completion time. In autonomous vehicle control, the cost function is related to the safety margin to obstacles. In contrast to traditional CPCSs, WHMC integrates human operators, wireless communication networks, and control systems into a tightly coupled framework characterized by strong dynamics and interdependencies. This integration results in a novel topology featuring intertwined wireless human and autonomous control loops. Consequently, WHMC involves three interconnected dynamic layers: 1) human state dynamics influencing the operator’s control decisions, 2) control system dynamics governing machine behavior, and 3) wireless channel dynamics affecting communication reliability and latency. Therefore, the performance metrics of WHMC are inherently more complex and less tractable than those of conventional CPCSs.

WHMC introduces a multitude of new performance metrics for its design and implementation. In WHMC, the quality of collaboration (QoC) is multi-goal-oriented and depends on diverse factors related to the three-level dynamics with the full consideration of task performance and human wellness. 
%Since WHMC involves a shared control task between humans and machines, task-level control performance (e.g., position accuracy and motion-tracking error) remains the main focus. However, as the human operator is an integral part of the system, factors related to human satisfaction, including user experience, preference, trust, fluency, and workload, should also be considered, as they may not always align with task-level performance metrics. For example, optimizing control accuracy by increasing computation resources may raise latency and then compromise human perceptual fluency, or vice versa. 
%Therefore, these performance metrics must be jointly optimized and be context-aware. Designing these objectives in isolation often leads to suboptimal QoC performance.
\gp{We link QoC to measurable metrics across the three coupled layers. At the task layer, QoC reflects control/mission outcomes (e.g., tracking error, completion time). At the human layer, QoC captures perceptual fluency and workload (e.g., delay/jitter-induced discontinuity, interaction smoothness, intervention frequency). At the network layer, QoC is primarily shaped by latency, jitter, reliability, and sustained capacity availability. For instance, immersive teleoperation is typically jitter-limited (fluency), while supervisory control is more outage- and availability-limited (safety and continuity).}

%\vspace{-0.3cm}
\subsection{Design Methodologies of WHMC}
\begin{figure}[!t]
\centering
\includegraphics[width=3.4in]{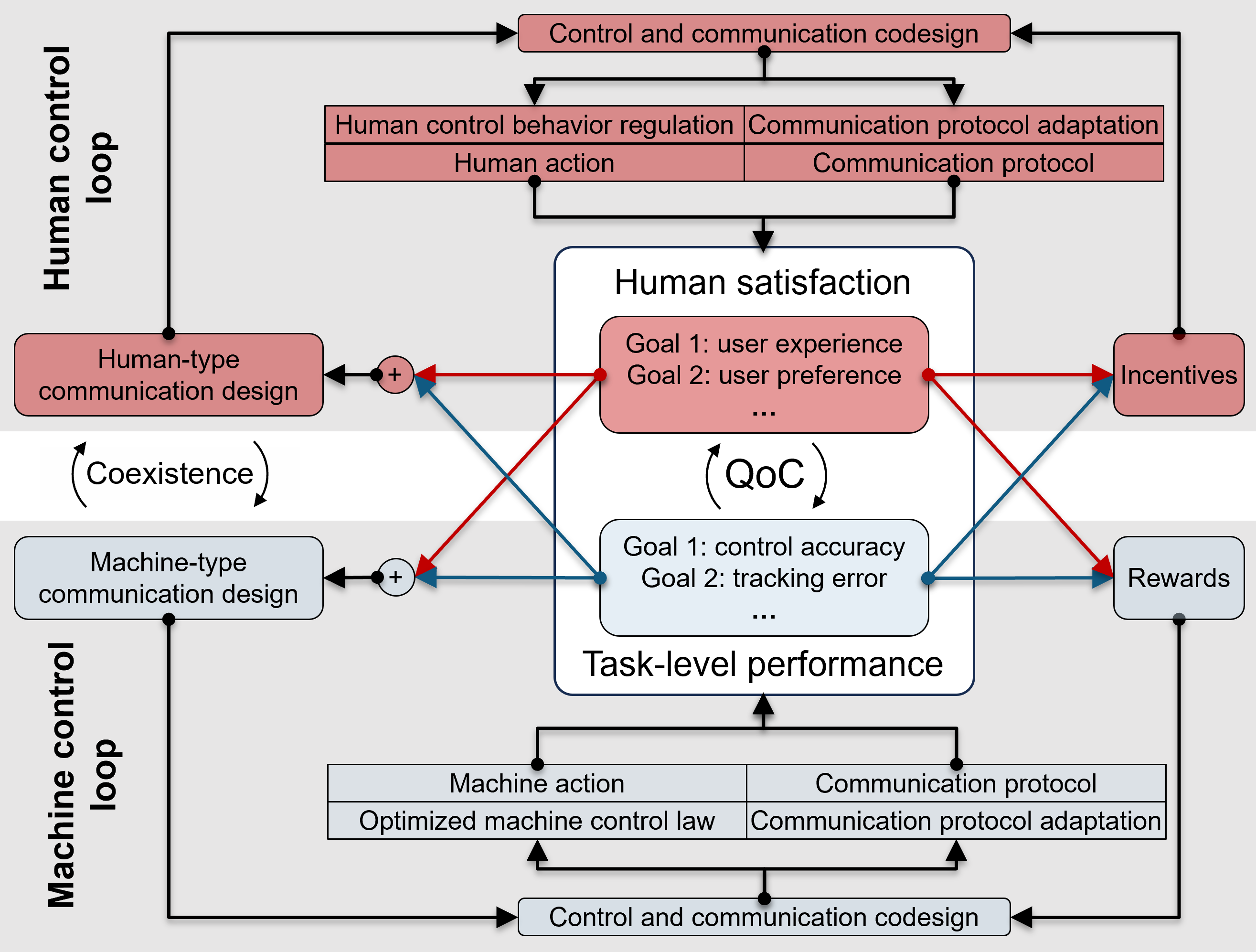}
%\vspace{-0.4cm}
\caption{The design methodology for WHMC, where multi-goal-oriented QoC bridges the fundamental design and the desired collaboration goals.}
\label{fig_6}
%\vspace{-0.0cm}
\end{figure}

Once QoC performance metrics are well established, new design methodologies are necessary to optimize WHMC. Human-type communications (HTC) and machine-type communications (MTC) will naturally coexist in WHMC. They have significantly different communication requirements, data behaviors, and data structures, which should be considered during design. Herein, we propose a WHMC optimization methodology, as shown in Fig.~\ref{fig_6}.

\subsubsection{QoC-aware codesign of HTC and MTC}
QoC-aware communications are essential to WHMC to guarantee application-level performance. Their design involves the coordination and synchronization of HTC and MTC for the human control loop and the machine control loop, respectively, which must be jointly optimized to guarantee QoC. Given the multi-goal-oriented nature of QoC, the distinct communication requirements of HTC and MTC are supported by tailoring wireless links, while mitigating their inter-link interference. The design of each link is adaptive to dynamic human control behaviors, collaborative control tasks, and time-varying wireless environments using historical data, such as channel/human states and human/machine control actions. The design also requires a comprehensive investigation of the relationship between QoC and the communication protocol, frame or packet structure, and transmission strategy. In addition, power and bandwidth allocation among multiple WHMC systems must reflect heterogeneous QoC-aware communication demands.

\subsubsection{QoC-oriented control-communication codesign}
Unlike traditional communication-oriented systems, a WHMC system emphasizes the goal-oriented codesign of communication and control systems with explicit awareness of human involvement. Most existing communication architectures are agnostic to collaborative control objectives. Their design focuses primarily on maximizing communication performance (e.g., throughput, reliability) rather than optimizing control performance or human satisfaction. 
Understanding collaborative control dynamics and human characteristics offers an untapped opportunity to relax the stringent communication requirements. 
The key is to prioritize transmission packets based on their contribution to QoC. This approach contrasts sharply with current systems (e.g., 5G), where all packets are treated with equal priority and only over-the-air transmission errors are considered.

\subsubsection{Feedback via reward shaping and incentive mechanism}
Optimizing WHMC performance requires instantaneous feedback that captures the synergistic interplay among humans, machine controllers, and wireless networks. When human control actions are not observable by machines, i.e., no trustworthiness in Fig.~\ref{fig_2}, the resulting human inputs can unintentionally disrupt machine operations.
To address this, the desired QoC objectives can be embedded into the joint design of these entities, enabling a unified feedback framework. QoC can be interpreted as instantaneous feedback signals (i.e., rewards for machines and incentives for humans) derived from the combined effects of wireless communication design, human control behavior, and machine control strategies on WHMC performance. Furthermore, WHMC naturally supports human–machine co-adaptation and co-evolution, allowing both agents to estimate and anticipate each other’s performance, thereby enhancing overall collaboration effectiveness.

%\vspace{-0.3cm}
\section{Design Requirements and Building Blocks}
\subsection{Key Communication Design Considerations}

Wireless communication, as a key entity, plays a pivotal role in supporting various WHMC applications. The detailed service requirements for CPCSs and for professional multimedia (video, imaging, and audio) applications have been extensively summarized in 3GPP TS 22.104 and 3GPP TS 22.263, respectively. However, defining service requirements for diverse WHMC applications is significantly more challenging.

Conventional 5G service categories, i.e., URLLC, enhanced mobile broadband (eMBB), massive machine-type communications (mMTC),  address low latency, high throughput, and massive connectivity separately. 
\gp{WHMC is generationally agnostic and can be instantiated on current 5G networks (e.g., 5G standalone with slicing) for near-term deployments, while benefiting from future 6G enhancements (e.g. high-rate sensing, AI-native slicing, and extended coverage) for broader WHMC applications.}
%However, WHMC demands demands all three simultaneously and at more extreme levels. This cannot be achieved by current 5G networks. For instance, delivering sub-5 ms latency with five-nines reliability with high data rates across wide-area links and dense deployments is far beyond 5G capabilities. 
Here, we identify five key considerations for WHMC communication. They form the fundamental communication requirements that enable effective WHMC.

\subsubsection{Accessibility}
Accessibility is the ability of a WHMC system to establish and maintain connectivity wherever humans and machines operate.
As human activity increasingly extends to space, aerial, maritime, and remote regions, WHMC is essential in seamless collaboration. Wireless communications are critical to bridging these physical distances and ensuring global connectivity.
Long-distance WHMC communications across different altitudes challenge 5G terrestrial networks, which cannot provide global connectivity. 
For instance, remote industrial or disaster-response missions may require connectivity spanning hundreds of kilometers via drones or satellites, far beyond terrestrial 5G capabilities.

\subsubsection{Transparency}
Transparency describes how well the human operator perceives the environment and the slaved machine as an extension of the self, and how well the slaved machine and human operator comprehend each other. Achieving this requires URLLC with low-jitter, high-fidelity and real-time communications, especially for geographically separated entities, whereby transmissions may traverse beyond the wireless access segment.
\gp{WHMC requires transport-aware URLLC. Flexible x-haul combined with slice-aware orchestration can reserve deterministic paths for control loops while scaling XR/multimedia streams.}
In practice, immersive mission-critical teleoperation (e.g., telesurgery or remote robot control) demands end-to-end latencies of a few milliseconds with jitter below 1 ms, exceeding current 5G URLLC performance.

\subsubsection{Scalability}
Scalability determines a WHMC system’s ability to accommodate vast numbers of agents and dynamically on-board new participants. Network topology may evolve rapidly as agents join or leave the collaboration. For instance, in a rescue operation involving human-assisted drone swarms through VR/AR interfaces, hundreds of drones may require simultaneous wireless control loops for collision avoidance and intra-swarm visual coordination, i.e., non-orthogonal human coordination control in Section~\ref{sec:agentMulti}.
Large-scale WHMC will demand wireless networks capable of integrating both high data rates and massive connectivity, going beyond 5G’s separately enhanced eMBB and mMTC categories.

\subsubsection{Resilience}
Resilience is vital for WHMC applications involving high-mobility agents, such as remote driving and autonomous fleet management, which impose stringent continuity and reliability requirements. Frequent handovers in high-speed scenarios can cause service interruptions, undermining control stability. For example, a 120 km/h vehicle may experience a handover every few seconds; even a millisecond-scale disruption can destabilize the control loop. Maintaining reliable URLLC during such transitions demands near-seamless mobility management, exceeding the resilience currently achievable with 5G networks.

\subsubsection{Sustainability}
Since many devices in WHMC, especially sensors in Internet of Things (IoT), wearable interfaces, mobile manipulators, and even satellites, are battery-powered, enhancing energy efficiency is important. Inefficient energy management and transmission rate control can lead to energy starvation, which may disable core components of WHMC in emergencies. 
For example, industrial IoT sensors may run for years on a single battery, and wearable HMIs often operate on milliwatt-level power budgets. Meeting these longevity requirements calls for ultra-energy-efficient communication strategies, which is a key focus of green 6G design.

%\vspace{-0.3cm}
\subsection{Promising Communication Technologies}\label{sec:6gTech}
The aforementioned gaps call for the development of advanced and deeply integrated communication technologies. Several emerging 6G technologies and multiple complementary technologies can be jointly applied to support WHMC.

\subsubsection{Human-bond communications and beyond}
Immersive user experiences enhance human control-oriented decision-making by providing realistic feedback, improving situational awareness, reducing cognitive load, and increasing engagement. Human bond communication (HBC) aspires to achieve this goal by integrating, digitizing, transmitting, and replicating the five human senses \cite{ref3}. Effective communication should also capture human well-being and functioning to build trust and foster reliable collaboration. This requires the communication of human control intentions, behavioral patterns, psychological states (e.g., emotions), physical states (e.g., fatigue), and physiological states (e.g., heart rate). Developing this communication system requires communication technologies beyond HBC, such as holographic communication \cite{ref4} and affective communication \cite{ref5}.

\subsubsection{Semantic and goal-oriented communication}
WHMC involves multi-modal human interaction, including text, speech, image, and body motion/gesture, requiring devices with similar interaction capabilities. This motivates the development of semantic communication systems, which can extract and transmit the semantic meaning of complex information rather than its raw bits in a noisy channel \cite{ref8}. Semantic communication relies primarily on a shared knowledge base that is understandable to both human operators and machines. Communication efficiency and reliability can be further improved by exploiting shared knowledge for semantic inference. Recent advances in AI technologies have boosted the potential of semantic communications in future 6G networks.

\subsubsection{Integrated sensing and communication}
The proliferation of wireless devices in large-scale WHMC can lead to severe spectrum congestion. Integrated sensing and communication (ISAC) is promising to address this issue~\cite{ref7}. ISAC allows hardware and spectrum sharing between sensing and communication functions, which reduces the overall hardware cost and improves spectrum efficiency. Empowered by ISAC, location-aware communication services, such as mobility management, become more accessible, increasing connectivity for WHMC. Combined with IoT, ISAC also supports human activity recognition, vital signal monitoring, and spatial-aware computing, advancing HMIs towards HBC and beyond.

\subsubsection{Other communication technologies}
Terahertz (THz) communication provides terabit-per-second (Tbps) throughput \cite{ref9} for data-intensive WHMC but is vulnerable to propagation environments. Reconfigurable intelligent surfaces (RISs) can shape reflections to control propagation environments \cite{ref1}. Besides, embedding RIS nodes in non-terrestrial networks forms space–air–ground–sea integrated networks (SAGSIN), providing flexible coverage \cite{ref11}. However, massive HTC–MTC coexistence in SAGSIN strains signaling, interference, and energy. Rate-splitting multiple access (RSMA) mitigates this by splitting common/private streams, enhancing spatial multiplexing, connectivity, efficiency, and reliability \cite{ref12}.

%\vspace{-0.3cm}
\subsection{Beyond Communication Technologies}\label{sec:OtherTech}
While communication technologies are crucial, they alone cannot enable WHMC systems. Advancing this paradigm demands attention to societal and privacy issues, as well as broader enabling technologies. Sensors, actuators, and HMIs serve as key interfacing technologies for seamless human–machine–environment interaction. Flexible, stretchable sensors made from soft materials can conform to the body, robots, or surroundings for continuous contextual monitoring \cite{ref13}. Collaborative robots (cobots) provide adaptable, portable, and cost-effective actuation for collaborative, teleoperated, or autonomous WHMC tasks. Next-generation HMIs, such as XR, motion-capture, tactile, and brain–computer interfaces \cite{ref5}, offer higher control freedom than conventional HMIs. 

Furthermore, WHMC relies on a programmable control plane that fuses communication and computation through SDN, NFV, and cloud–edge–end orchestration \cite{ref14}. 
\gp{In current deployments, network slicing in the 5G standalone core can provide traffic isolation between (i) latency-critical control loops and (ii) high-rate sensing streams, while cross-domain orchestration binds RAN, transport, and edge resources into a per-task service chain. In addition, an O-RAN-style radio intelligent controller illustrates how near-real-time RAN policies (e.g., scheduling) can be adapted for QoC-aware operation. Adaptive functional splits provide a practical knob to trade QoC vs. latency by shifting processing closer to the radio~\cite{ref16}.}
%Intent-driven interfaces compose per-task service chains and enforce isolation, scaling, and governance across domains, enabling consistent world models and semantics for WHMC. Digital twins replicate WHMC systems in cyberspace \cite{ref15}, providing immersive interfaces and live testbeds for reprogramming complex operations; however, end-device, rendering, and backhaul latency still hinder real-time performance, requiring synergy among interfacing, computing, and 6G networking.

Finally, effective control strategies remain the backbone of WHMC, which ensures safe and efficient cooperation under network imperfections. Shared autonomy framework allocates control authority based on confidence, workload, and trust; haptic control and teleportation maintain stability via passivity-based compensation and predictive feedback; communication-aware control (e.g., event/self-triggered and anytime control) prioritizes transmissions by criticality; and learning-enabled control (e.g., safe reinforcement learning) adapts to environmental and human uncertainties. Integrating these control, computing, interfacing, and digital-twin technologies ensures that WHMC systems remain stable, responsive, and efficient.

\gp{
\subsection{Deployment Roadmap}
In the near term, WHMC is feasible in factory settings using private 5G/5G-Advanced with edge computing and traffic isolation for control loops. Early 6G deployments are expected to broaden WHMC by improving end-to-end latency/jitter control, integrating AI-assisted orchestration, and enabling richer XR-based interfaces. Later 6G evolution can extend WHMC to wide-area and multi-altitude scenarios by coupling non-terrestrial connectivity with transport-aware slicing across RAN and optical x-haul. 
%The key expectation of 6G is to make WHMC systems scalable and robust across sites.

From a maturity perspective, slicing and edge-enabled service chaining are available in 5G standalone systems and can be used to separate control-loop traffic from perception streams. In early 6G, tighter and more automated cross-domain orchestration, improved deterministic performance, and better support for advanced XR are expected to become mainstream. Technologies, such as semantic/goal-oriented communication, ISAC, and THz/RIS-based ubiquitous coverage, remain sensitive to environments and standardization uncertainty, and are therefore positioned as longer-term accelerators.
}

%\vspace{-0.3cm}
\section{Illustrative Case Study}
\subsection{Experiment Setup}
\begin{figure}[!t]
\centering
%\color{blue}
\includegraphics[width=3.4in]{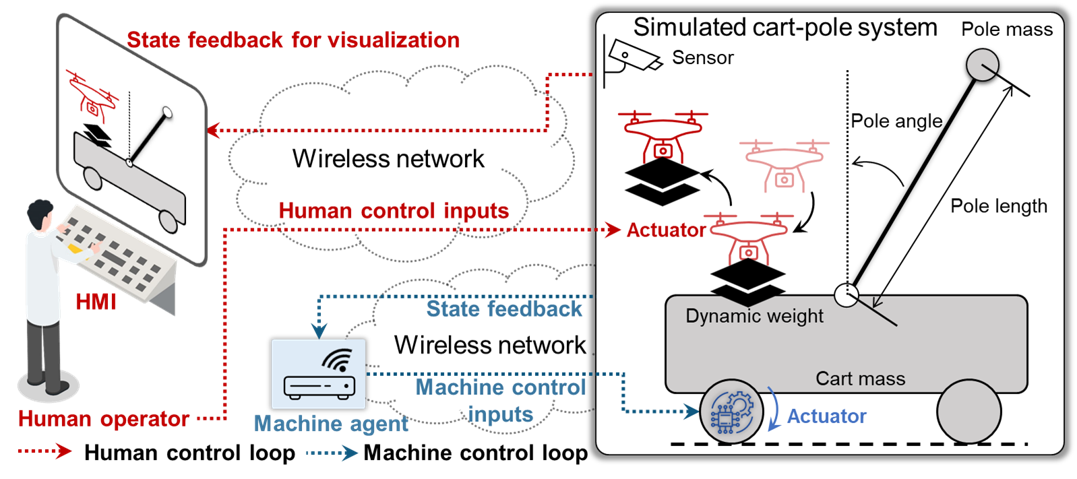}
%\vspace{-0.4cm}
\caption{The wireless cart-pole system for the case study of WHMC.}
\label{fig_4}
%\vspace{-0.1cm}
\end{figure}

Motivated by typical manufacturing scenarios, we examine a wireless cart-pole control system, as shown in Fig.~\ref{fig_4}. It is a classic nonlinear and unstable control problem. The machine linearizes the cart-pole dynamics around the equilibrium point and uses a standard linear quadratic regulator to apply forces to the cart and balance the pole. A challenging scenario is introduced by adding a dynamic weight to the cart, which the machine cannot detect or compensate for. However, a human operator, who can observe this disturbance, intervenes by pressing a key to remove the weight, preventing system failure. 
The cart-pole system, machine controller, and human operator are spatially separated, communicating over a wireless network. Control signals and system states are exchanged using short-packet transmissions to ensure low latency.
The average channel gains follow the free-space path loss model. The time-varying channel gains are generated from Rayleigh fading channel models with a unit mean. Communication parameters are summarized in Table~\ref{tab:table1}. The mass of the cart, pole, and weight are 10 kg, 4 kg, and 5 kg, respectively. The pole length is 4 m. The initial pole angle is $\pi$/6.

%\vspace{-0.3cm}
\subsection{Results and Discussion}
\begin{figure}[!t]
\centering
%\color{blue}
\includegraphics[width=3.4in]{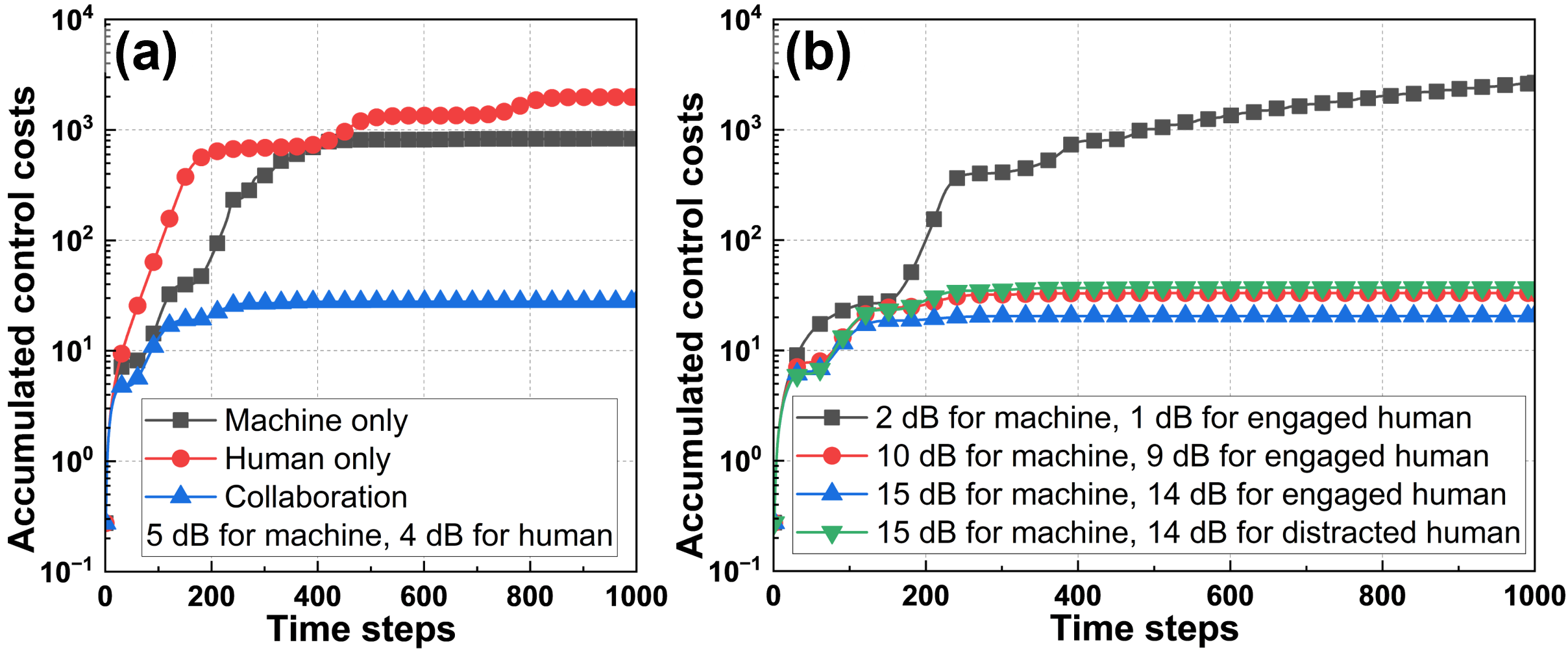}
%\vspace{-0.4cm}
\caption{Evaluation of WHMC: (a) Impacts of decision-makers on the control performance. (b) Impacts of communications on the control performance.}
\label{fig_5}
%\vspace{-0.1cm}
\end{figure}

The control performance is evaluated at each time step using a quadratic cost function on the deviation of the pole angle, reflecting the system’s ability to maintain balance.
A lower cost indicates better control. 
Fig.~\ref{fig_5}(a) compares the accumulated control cost across three scenarios: machine-only control, human-only control, and WHMC (i.e., a machine-dominated system in Fig.~\ref{fig_2}). The WHMC scenario shows a slower increase in cost and reaches a significantly lower steady-state value, demonstrating that collaborative control yields superior performance. 
Fig.~\ref{fig_5}(b) presents the impact of signal-to-noise ratio (SNR) on control performance in the WHMC case.
When the human operator is actively engaged, higher SNRs lead to improved control outcomes. However, when the operator is distracted (e.g., due to fatigue), increasing SNR does not enhance performance. 
%This highlights that communication improvements alone are insufficient; effective collaboration requires integrated optimization across control algorithms, communication reliability, and human engagement. 
\gp{This case study is a minimal proxy for WHMC, demonstrating the complementarity of human resilience and machine speed. The autonomous controller stabilizes fast dynamics, while the human operator provides sparse interventions when unmodeled disturbances occur. Communication quality affects performance. Effective collaboration requires integrated optimization across control, communication, and human engagement, motivating the definition of QoC to jointly capture control outcomes, network key performance indicators (KPIs), and human state.}

\begin{table}[!t]
\footnotesize
\setlength\tabcolsep{1.0pt}
\caption{Summary of Communication Parameters of the Case Study\label{tab:table1}}
%\vspace{-0.3cm}
\centering
\begin{tabular}{ll|ll}
\hline\hline
\textbf{Communication parameters}            & \textbf{Value}  & \textbf{Free-space path loss model}           & \textbf{Value}    \\ \hline
Code rate {[}bps{]}                 & 2      & Antenna gain                         & 4        \\
Packet length {[}symbols{]}         & 500    & Carrier frequency {[}MHz{]}          & 915      \\
Min transmit power {[}dBm{]}        & 20     & Distances to plant {[}m{]}    & 45-50       \\
Background noise power {[}dBm{]}    & -70    & Path loss exponent                   & 2.9       \\\hline\hline
\end{tabular}
%\vspace{-0.0cm}
\end{table}

%\vspace{-0.3cm}
\section{Open Challenges and Future Directions}
\subsection{Fundamental Mathematical Modeling}
At present, there is a lack of robust theories and tools for designing flexible and scalable WHMC systems. Unlike conventional control systems in structured and repetitive settings, WHMC systems involve dynamic interactions between human and machine via wireless communication. This requires fundamentally new mathematical models and analytical frameworks that can manage system complexity, capture essential collaborative human–machine dynamics, and remain mathematically tractable. Future research must focus on collaborative task abstraction, human behavior modeling, and wireless channel characterization in complex environments.

\subsection{Tractable Performance and Stability Analysis}
There is a mismatch between traditional communication KPIs and the QoC requirements in WHMC. Packets differ in importance, deadlines, and reliability, so KPI gains alone may waste resources without improving QoC. A tractable framework must jointly capture QoC and packet-level dynamics while linking them to control outcomes. Equally critical is stability. Transmission errors and delays can destabilize closed loops, yet formalizing stability is hard without faithful human behavior models. A unified framework that captures diverse performance goals and human effort remains elusive. Progress demands an interdisciplinary approach to characterize fundamental limits of WHMC.

%\vspace{-0.3cm}
\subsection{Optimal Design of Large-Scale Systems}
The first step toward optimal design of large-scale WHMC systems is to formulate problems that account for the three-layer dynamics—control, communication, and human behavior. However, deriving closed-form expressions for objective functions and stability constraints is challenging due to the analytical intractability of many performance metrics, leading to NP-hard mixed-integer nonlinear programs. Yet, current AI-based approaches typically treat machine controllers, human operators, and wireless networks in isolation, limiting their effectiveness. Developing integrated AI frameworks for diverse WHMC applications is vital for scalability through improved model adaptation, inference efficiency, and interpretability.

%\vspace{-0.3cm}
\subsection{Critical Issues beyond Technology} 
WHMC also brings critical considerations beyond technology. The deployment of wireless communication systems is closely tied to upstream industries, which may lag behind the rapid advancements in communication theories. Ethical and societal considerations are increasingly relevant (e.g., potential health risks from dense network deployment and high-frequency transmissions), as well as privacy and surveillance issues in a sensor-rich environment. It is crucial to balance technological innovations with socio-economic factors to ensure global accessibility, affordability, and trust.

\section{Conclusion}
This paper presents a forward-looking vision for wireless-empowered HMC in the 6G era. 
We highlighted the distinct performance metrics of WHMC compared to traditional communication systems, emphasizing the importance of human satisfaction and task-level outcomes. A design methodology tailored to WHMC has been proposed, recognizing the need to integrate knowledge from control, communication, and human factors. Our proof-of-concept study demonstrates the tangible benefits of WHMC, and we have outlined key future directions, including scalable system design, tractable performance and stability analysis, and ethical deployment.

%\vspace{-0.3cm}

%\newpage

%\section*{Biographies}
\vspace{-33pt}

\begin{IEEEbiographynophoto}{Gaoyang Pang}
[M] is a Post-Doctoral Research Associate with the School of Electrical and Computer Engineering (ECE), The University of Sydney (USyd), Australia.
\end{IEEEbiographynophoto}

\vspace{-33pt}

\begin{IEEEbiographynophoto}{Wanchun Liu}
[SM] is a Senior Lecture and an ARC DECRA Fellow at USyd. Her research interests are networked robotics, industrial IoT, and HMC. 
\end{IEEEbiographynophoto}

\vspace{-33pt}

\begin{IEEEbiographynophoto}{Chentao Yue}
[M] is an ARC DECRA Fellow at USyd. His research interests are in the areas of coding theory and semantic communications.
\end{IEEEbiographynophoto}

%\vspace{-33pt}

%\begin{IEEEbiographynophoto}{Dusit Niyato}
%[F] is a professor in the College of Computing and Data Science, at Nanyang Technological University, Singapore. His research interests are in the areas of sustainability, edge intelligence, decentralized machine learning, and incentive mechanism design.

\vspace{-33pt}

\begin{IEEEbiographynophoto}{Daniel E. Quevedo}
[F] is a Professor of cyber-physical systems at the ECE, USyd. His research interests are in networked control systems, control of power converters, and cyber-physical systems security.
\end{IEEEbiographynophoto}

\vspace{-33pt}

\begin{IEEEbiographynophoto}{Karl H. Johansson}
[F] is Swedish Research Council Distinguished Professor in electrical engineering and computer science with KTH Royal Institute of Technology in Sweden and Founding Director of Digital Futures. 
\end{IEEEbiographynophoto}

\vspace{-33pt}

\begin{IEEEbiographynophoto}{Branka Vucetic}
[LF] is an Australian Laureate Fellow, a Professor of Telecommunications, and Director of the Centre for IoT and Telecommunications at USyd. She is a Fellow of the Australian Academy of Technological Sciences and Engineering and the Australian Academy of Science. 
\end{IEEEbiographynophoto}

\vspace{-33pt}

\begin{IEEEbiographynophoto}{Yonghui Li}
[F] is an ARC Industry Laureate Fellow, a Professor of Telecommunications, and Director of Wireless Engineering Laboratory at the ECE, USyd. His research interests are millimeter wave communications, machine-to-machine communications, and coding techniques.
\end{IEEEbiographynophoto}
\vfill


\begin{thebibliography}{1}
\bibliographystyle{IEEEtran}

\bibitem{ref1}
%I. Kardush, S. Kim and E. Wong, “A techno-economic study of Industry 5.0 enterprise deployments for human-to-machine communications,” \textit{IEEE Commun. Mag.}, vol. 60, no. 12, pp. 74--80, 2022.
M. Noor-A-Rahim et al., “Toward Industry 5.0: Intelligent reflecting surface in smart manufacturing,” \textit{IEEE Commun. Mag.}, vol. 60, no. 10, pp. 72--78, 2022.

\bibitem{ref0}
\gp{
R. Sabella et al., “Robotics and industrial automation enabled by 5G - Ericsson,” \textit{Ericsson Technol. Rev.}, no. 2, pp. 1--12, 2018.
}

\bibitem{ref2}
C. -X. Wang et al., “On the road to 6G: Visions, requirements, key technologies, and testbeds,” \textit{IEEE Commun. Surveys Tuts.}, vol. 25, no. 2, pp. 905--974, 2023.

\bibitem{ref3_1}
G. Pang, W. Liu, D. Niyato, D. Quevedo, B. Vucetic and Y. Li, “Wireless human-machine collaboration in Industry 5.0,” \textit{IEEE Trans. Wirel. Commun.}, accepted, 2025.

\bibitem{ref3}
T. Iftikhar, H. A. Khattak, Z. Ameer, M. A. Shah, F. F. Qureshi and M. Z. Shakir, “Human bond communications: Architectures, challenges, and possibilities,” \textit{IEEE Commun. Mag.}, vol. 57, no. 2, pp. 19--25, 2019.


\bibitem{ref4}
T. Gong et al., “Holographic MIMO communications: Theoretical foundations, enabling technologies, and future directions,” \textit{IEEE Commun. Surveys Tuts.}, vol. 26, no. 1, pp. 196--257, 2023.

\bibitem{ref5}
D. Wu, B. -L. Lu, B. Hu and Z. Zeng, “Affective brain–computer interfaces (aBCIs): A tutorial,” \textit{Proc. IEEE}, vol. 111, no. 10, pp. 1314--1332, 2023.

% \bibitem{ref6}
% Q. Zhao, G. Li, J. Cai, M. Zhou and L. Feng, “A tutorial on Internet of Behaviors: Concept, architecture, technology, applications, and challenges,” \textit{IEEE Commun. Surveys Tuts.}, vol. 25, no. 2, pp. 1227--1260, 2023.

\bibitem{ref8}
W. Yang et al., “Semantic communications for future Internet: Fundamentals, applications, and challenges,” \textit{IEEE Commun. Surveys Tuts.}, vol. 25, no. 1, pp. 213--250, 2022.

\bibitem{ref7}
F. Liu et al., “Integrated sensing and communications: Toward dual-functional wireless networks for 6G and beyond,” \textit{IEEE J. Sel. Areas Commun.}, vol. 40, no. 6, pp. 1728--1767, 2022.

\bibitem{ref9}
Z. Chen et al., “Terahertz wireless communications for 2030 and beyond: A cutting-edge frontier,” \textit{IEEE Commun. Mag.}, vol. 59, no. 11, pp. 66--72, 2021.

% \bibitem{ref10}
% M. Noor-A-Rahim et al., “Toward Industry 5.0: Intelligent reflecting surface in smart manufacturing,” \textit{IEEE Commun. Mag.}, vol. 60, no. 10, pp. 72--78, 2022.

\bibitem{ref11}
M. M. Azari et al., “Evolution of non-terrestrial networks from 5G to 6G: A survey,” \textit{IEEE Commun. Surveys Tuts.}, vol. 24, no. 4, pp. 2633--2672, 2022.

\bibitem{ref12}
B. Clerckx et al., “A primer on rate-splitting multiple access: Tutorial, myths, and frequently asked questions,” \textit{IEEE J. Sel. Areas Commun.}, vol. 41, no. 5, pp. 1265--1308, 2023.

\bibitem{ref13}
G. Pang, G. Yang and Z. Pang, “Review of robot skin: A potential enabler for safe collaboration, immersive teleoperation, and affective interaction of future collaborative robots,” \textit{IEEE Trans. Med. Rob. Bio.}, vol. 3, no. 3, pp. 681--700, Aug. 2021.

\bibitem{ref14}
S. Duan et al., “Distributed artificial intelligence empowered by end-edge-cloud computing: A survey,” \textit{IEEE Commun. Surveys Tuts.}, vol. 25, no. 1, pp. 591--624, 2022.

% \bibitem{ref15}
% S. Mihai et al., “Digital twins: A survey on enabling technologies, challenges, trends and future prospects,” \textit{IEEE Commun. Surveys Tuts.}, vol. 24, no. 4, pp. 2255--2291, 2022.

\bibitem{ref16}
\gp{
I. Tomkos et al., "The “X-Factor” of 6G networks: Optical transport empowering 6G innovations," \textit{IT Prof.}, vol. 26, no. 1, pp. 32--39, 2024.
}

\end{thebibliography}
\end{document}